\pdfoutput=1
\documentclass[letterpaper, 10 pt, conference]{ieeeconf}  % Comment this line out if you need a4paper

\IEEEoverridecommandlockouts                              % This command is only needed if 
\usepackage{graphicx}
\usepackage{amsmath}
\usepackage{amssymb}
\usepackage{xcolor}
\usepackage{url}
\usepackage{booktabs} % For better table lines
\usepackage{tabularray} 
\usepackage{tabularx} % for 'tabularx' env. and 'X' col. type
\usepackage{ragged2e} % for \RaggedRight macro
\usepackage{booktabs} % for \toprule, \midrule etc macros
\usepackage{cuted}
\usepackage{array}      % 用于定义列的格式
\usepackage{multirow}   % 用于合并行
\usepackage{booktabs}   % 用于创建漂亮的表格线
\usepackage[utf8]{inputenc}
\usepackage{svg} 
\usepackage[ruled]{algorithm2e}
\usepackage{amsmath} 
\usepackage{graphicx}
\usepackage{amsmath}
\usepackage{amssymb}
\usepackage{mathrsfs} %
\usepackage{xcolor}
\usepackage{url}
\usepackage{caption}
\usepackage{booktabs} % For better table lines
\usepackage{tabularray} 
\usepackage{tabularx} % for 'tabularx' env. and 'X' col. type
\usepackage{ragged2e} % for \RaggedRight macro
\usepackage{booktabs} % for \toprule, \midrule etc macros
\usepackage{cuted}
\usepackage{array}      % 
\usepackage{multirow}   % 
\usepackage{booktabs}   %
\usepackage[utf8]{inputenc}
\usepackage{svg} 
\usepackage{amssymb}
\usepackage{algpseudocode}  
\usepackage{amsmath}  
\usepackage{booktabs}
\usepackage[colorlinks,linkcolor=blue]{hyperref}

\title{\LARGE \bf

Structural Optimization of Lightweight Bipedal Robot via SERL
}
\author{Yi Cheng*$^{1}$, Chenxi Han*$^{1}$, Yuheng Min$^{1}$, Linqi Ye†$^{2}$, Houde Liu†$^{1}$, Hang Liu$^{3}$\\\href{SERL-IROS2024.github.io}{SERL-IROS2024.github.io}
\thanks{* These authors contributed equally to this work.}
\thanks{† corresponding author.}
\thanks{Research supported by the National Natural Science Foundation of China under grants No.92248304 and Shenzhen Science Fund for Distinguished
Young Scholars under Grant RCJC20210706091946001}% <-
\thanks{$^{1}$ Tsinghua University, 100084 Beijing, China }%
\thanks{$^{2} $ Shanghai University, 200444 Shanghai, China. }
\thanks{$^{3} $ University of Michigan, Ann Arbor, MI 48109, USA}
}

\begin{document}

\let\oldtwocolumn\twocolumn

\renewcommand\twocolumn[1][]{%
    \oldtwocolumn[{#1}{
    \begin{flushleft}
        \centering
        \vspace{-30pt}
           
        \includegraphics[clip,trim=0cm 0cm 0cm 0cm,width=0.98\textwidth]{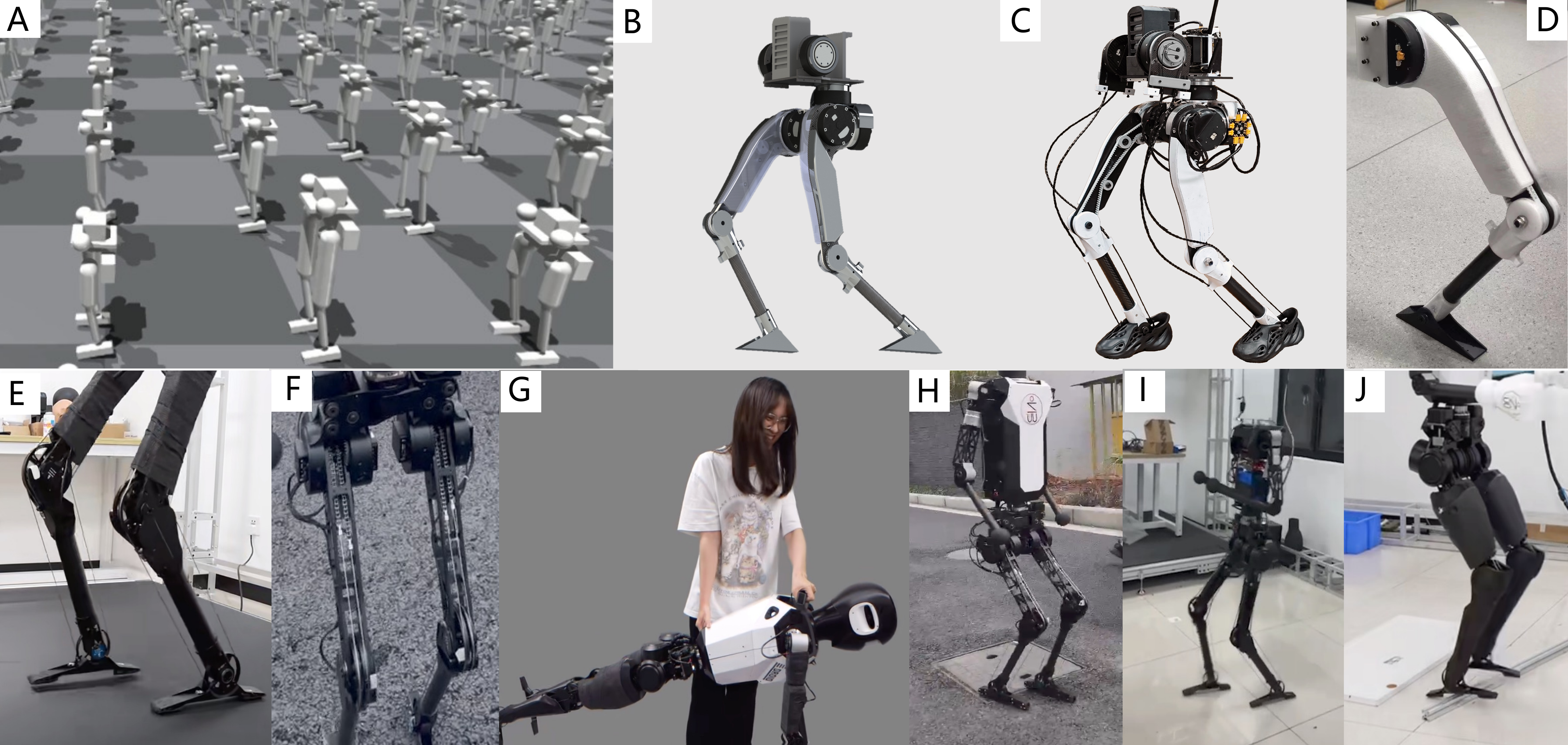}
        % \vspace{-20pt}
        
        \captionsetup{justification=justified} % Add this line
        
        \captionof{figure}{
             In Figures A-D, we utilized the SERL algorithm to design a lightweight bipedal robot from scratch, which we named Wow Orin. Through a series of experiments, we have demonstrated the effectiveness of the SERL algorithm and the outstanding performance of Wow Orin. A: Optimization process of the SERL algorithm in a simulation environment. B: Simulation model of Wow Orin designed based on the optimization results from the SERL algorithm. C: Physical Wow Orin manufactured and assembled based on the simulation model. Following the completion of the Wow Orin, enhancements were made to the leg structure's lightweight design, supplemented by the application of the SERL algorithm, leading to the development of the X02-LITE robot. Figures E-G depict its structural design, while Figures H-J illustrate the effectiveness of its locomotion capabilities. For further details, please see \href{SERL-IROS2024.github.io}{SERL-IROS2024.github.io}. The above aspects also underscore the significance and value of the research conducted. Note that all designs and experimental comparisons in this paper are based on the Wow Orin robot. 
        }\label{fig:setup}
    \end{flushleft}
    }]
}

\maketitle
\thispagestyle{empty}
\pagestyle{empty}

%%%%%%%%%%%%%%%%%%%%%%%%%%%%%%%%%%%%%%%%%%%%%%%%%%%%%%%%%%%%%%%%%%%%%%%%%%%%%%%%
\begin{abstract}

Designing a bipedal robot is a complex and challenging task, especially when dealing with a multitude of structural parameters. Traditional design methods often rely on human intuition and experience. However, such approaches are time-consuming, labor-intensive, lack theoretical guidance and hard to obtain optimal design results within vast design spaces, thus failing to full exploit the inherent performance potential of robots. In this context, this paper introduces the SERL (Structure Evolution Reinforcement Learning) algorithm, which combines reinforcement learning for locomotion tasks with evolution algorithms. The aim is to identify the optimal parameter combinations within a given multidimensional design space. Through the SERL algorithm, we successfully designed a bipedal robot named Wow Orin, where the optimal leg length are obtained through optimization based on body structure and motor torque. We have experimentally validated the effectiveness of the SERL algorithm, which is capable of optimizing the best structure within specified design space and task conditions. Additionally, to assess the performance gap between our designed robot and the current state-of-the-art robots, we compared Wow Orin with mainstream bipedal robots Cassie and Unitree H1. A series of experimental results demonstrate the Outstanding energy efficiency and performance of Wow Orin, further validating the feasibility of applying the SERL algorithm to practical design.

\end{abstract}

%%%%%%%%%%%%%%%%%%%%%%%%%%%%%%%%%%%%%%%%%%%%%%%%%%%%%%%%%%%%%%%%%%%%%%%%%%%%%%%%
\section{INTRODUCTION}

Designing a bipedal robot is a vast and intricate engineering task, primarily due to numerous parameters that require manual selection. These parameters encompass structural length, body weight, rotational inertia, among others, playing a crucial role in the motion performance and overall functionality of bipedal robots. However, researchers have long faced the challenge of making appropriate choices among numerous possible parameter combinations to design an outstanding bipedal robot.

In many cases, these parameters are subjectively chosen based on engineers' personal experiences. This empirical approach often demands engineers to possess extensive experience and undergo numerous trials and modifications to achieve satisfactory results. 
Additionally, some design methodologies have drawn inspiration from the structural characteristics of natural bipedal organisms\cite{c1}, adopting bio-inspired method to attempt replication of these organisms' structural parameters within robotic designs. However, considering the essential differences in the overall structure between robots and their biological models, such approaches often fail to fully exploit the potential performance aspects of the robot's hardware.

% \begin{figure}
%     \centering
%     \includegraphics[width=\columnwidth]{coverpage.pdf}    
%     \caption{In reactions A-D, we utilized the SERL algorithm to design a lightweight bipedal robot from scratch, which we named Wow Orin. Through a series of experiments, we have demonstrated the effectiveness of the SERL algorithm and the outstanding performance of Wow Orin. A: Optimization process of the SERL algorithm in a simulation environment. B: Simulation model of Wow Orin designed based on the optimization results from the SERL algorithm. C: Physical Wow Orin manufactured and assembled based on the simulation model. Following the completion of the Wow Orin, enhancements were made to the leg structure's lightweight design, supplemented by the application of the SERL algorithm, leading to the development of the Xiao Bei robot.  This robot boasts an exceptionally lightweight frame, with a height of 1.6 meters and a total mass of merely 28 kg.  Figures E-F depict its structural design, while Figures F-I illustrate the effectiveness of its locomotion capabilities.  For further details, please see...  The above aspects also underscore the significance and value of the research conducted. Note that all designs and experimental comparisons in this paper are based on the wow Orin robot. }
%     \label{fig:example}
% \end{figure}

Furthermore, current research tends to overly emphasize either the control domain or the singular aspect of structural design, neglecting the inseparable relationship between the structure and control policy of bipedal robots. Isolated studies on either structural design or control policy may not yield optimal results. To address this research gap, collaborative design methods have emerged in recent years, treating structural design and control policy as an integral whole. However, most of these methods are currently predominantly applied to the design and simulation testing of soft-bodied robots, with relatively less application in the practical design of rigid-bodied robots.

To tackle these challenges, this paper proposes an innovative algorithm known as the Structure Evolution Reinforcement Learning Algorithm (SERL). This algorithm can end-to-end generate optimized design structures under given task goals and design requirements, while concurrently producing robust task completion policy. SERL algorithm achieves this by training the control policy of bipedal robots through reinforcement learning and implementing self-evolution of robot structural parameters via genetic algorithms. In comparison to methods relying on manual adjustments and bio-inspiration, the SERL algorithm significantly enhances design efficiency, reduces workload, and ensures the rational reliability of design results.

Then we developed a bipedal robot with nine degrees of freedom, named Wow Orin. The design process initially involves defining the necessary motor parameters and transmission methods. We innovatively adopted a method utilizing innovative fishbone-barb line actuation and belt drive, which relocates the drive motors for the ankle and knee joints to the top of the robot, reducing the rotational inertia of leg movements. Simultaneously, we employed the SERL algorithm to optimize the leg length parameters within the specified design space, obtaining the optimal leg length for comprehensive motion tasks. We compared the optimized robot model with a manually specified leg length parameter model, confirming that the SERL algorithm optimized model performs significantly better in meeting task requirements. Furthermore, we conducted a comprehensive comparison between "Wow Orin" and advanced bipedal robots Cassie and Unitree H1, demonstrating the outstanding performance of Wow Orin in terms of energy efficiency and agility.
The main contributions of this paper include:
\begin{itemize}
\item Introduce a novel bipedal robot structural parameter design method—the SERL algorithm, capable of finding the optimal parameter combination for specific tasks within a predefined design space.
\item Propose a two-stage optimization training framework, achieving integrated optimization of structural parameters and motion control policy.
\item Using the optimized structural parameters from the SERL algorithm, combined with innovative fishbone-barb line actuation and belt drive, we designed a flexible and lightweight bipedal robot, demonstrating its significant advantages over existing bipedal robots.
\end{itemize}

%%%%%%%%%%%%%%%%%%%%%%%%%%%%%%%%%%%%%%%%%%%%%%%%%%%%%%%%%%%%%%%%%%%%%%%%%%%%%%%%
\section{RELATER WORK}

\subsection{Evolution Algorithms for Robot Structure Design}

Evolution policy is a type of optimization algorithm inspired by the principles of evolution in the natural world, primarily employed to address optimization problems in continuous parameter spaces. Evolution policy possesses a broader capability to explore design spaces, especially adept at handling complex and multivariate design challenges. Researchers have successfully integrated evolution policy into the co-design of robots with simplified models for simple tasks, achieving commendable results\cite{c3,c4,c5}.
In addition,  \cite{c5}, \cite{c7}, and \cite{c8} share a common emphasis on combining evolutionary strategies with reinforcement learning techniques to address various challenges. Specifically, they focus on evolving agent morphologies and optimizing robot designs to enhance locomotion control and adaptability in complex environments. For instance, \cite{c5} introduces the Deep Evolutionary Reinforcement Learning (DERL) algorithm, which enables the evolution of diverse agent morphologies using low-level sensory inputs. Similarly, \cite{c7} utilizes meta reinforcement learning and genetic algorithms to optimize legged robot designs, achieving improved adaptability across different environmental conditions. Additionally, \cite{c8} presents Evolution Gym, a benchmark platform for co-optimizing soft robot design and control through co-evolution algorithms.

It is worth noting that existing research methods often focus on the design of soft robots or robots in virtual environments. However, we focus on the rigid bipedal robot and validates the designs through physical experiments, showcasing the feasibility of our optimization approach in more realistic scenarios.

\subsection{RL for Locomotion Control of Legged Robots}

The locomotion control of legged robots is a complex engineering task. Traditional methods involve the physical dynamic modeling of legged robots and the application of control theory to address this challenge. However, such an approach requires designers to possess extensive expertise and practical experience.

In recent years, reinforcement learning methods have shown tremendous potential in the mobility tasks of legged robots, with many studies significantly enhancing the mobility robustness and agility of legged robots through the use of reinforcement learning. These methods have been particularly applied to quadruped robots, achieving robust outdoor walking, high-speed running \cite{Quadrupeds Running1}, and stair climbing \cite{c10, c12,Quadrupeds stair4}, among other remarkable feats such as parkour \cite{Quadrupeds parkour1,Quadrupeds parkour2}. Compared to quadrupeds, the design of motion controllers for bipedal robots is more challenging. Current approaches either optimize through reward design and feature engineering or incorporate prior reference motions (trajectories, gait parameters, etc.) to achieve robust walking, jumping, and running tasks for bipedal robots \cite{Biped rubust walk1,Biped rubust walk2,Biped rubust walk3}. These efforts typically involve designing specialized controllers for robots after their physical systems have been fully developed. 

Unlike the aforementioned works, our approach begins from the structural design phase, considering the compatibility between structure and motion control, and proposes a hybrid optimization method that simultaneously optimizes structural design and learns optimal control.

\subsection{Co-Design of Legged Robots}
The parameter design of robots is often deeply associated with specific tasks, particularly for complex nonlinear systems like legged robots. Structural parameters significantly impact the performance in various tasks. To achieve the structural design of legged robots for specific tasks, the scientific community has explored and attempted various methods. For instance, \cite{Humanoid Robot Co-design} uses the HZD gait generation framework to optimize for stable gait and leg length. \cite{Codesign of Humanoid}, \cite{Computational design}, \cite{Versatile Co-design}seek the optimal robot design through a bilevel optimization method. Additionally, other works such as \cite{Large-scale admm-based co-design}, \cite{Computational co-optimization}, \cite{One robot for many tasks} employ trajectory optimization and ADMM methods for co-design.

Unlike these methods, our optimization procedure is based on reinforcement learning approach that does not require cumbersome task modeling and algorithm parameter design, making it more generalizable to different tasks. This approach simplifies the design process and allows for greater flexibility in adapting to various task requirements.

\section{Method}
%*********************************************************************

\subsection{Structural Evolution Reinforcement Learning} 

\subsubsection{Overview}
In response to the black-box optimization problem regarding the impact of structural parameters on the comprehensive motion performance of biped robots, this study introduces a Structural Evolution Reinforcement Learning algorithm (SERL). SERL integrates the dynamic adjustment capabilities of reinforcement learning policy with the global search capacity of genetic algorithms. Without reliance on explicit model guidance, it explores the design parameter space (specified in this paper as the leg length of the bipedal robot), with the aim of maximizing task rewards and optimizing the best size design and control policy. This method effectively combines the strengths of both algorithms, not only finding the optimal structural parameters but also developing a highly robust motion control policy. 
Due to the limited information acquired from the robot's onboard sensors, training directly within the body's proprioceptive observation space makes it challenging to uncover the bipedal robot's limit performance. Therefore, we propose the implementation of a two-phase training architecture, we introduce additional observations in the first stage of training and distill them into a deployable policy in the second stage.  Figure 2 and Figure 3  present the details.

% In order to achieve the optimal design of bipedal robot structures, this paper proposes the SERL algorithm, a method that combines reinforcement learning with genetic evolution algorithms. The algorithm aims to maximize rewards and optimize the best size design and control policy within a given design space (specified in this paper as the leg length of the bipedal robot). Figure 2 illustrates the overall framework of the SERL algorithm.

\begin{figure}[h]
    \centering
    \includegraphics[width=\columnwidth]{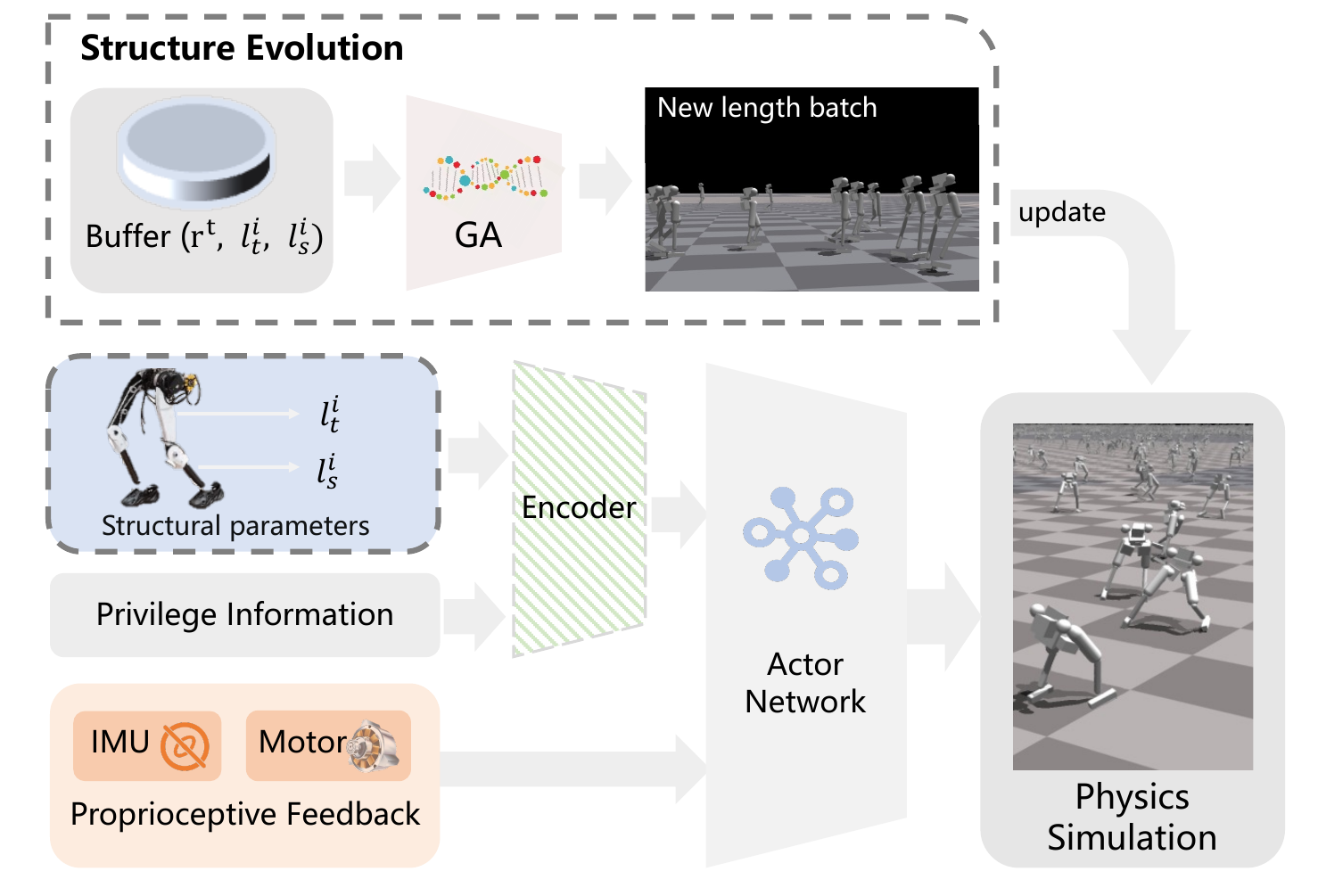}
    \caption{Overview of SERL framework. We employ a two-stage training method. Figure 2 illustrates the first stage, the SERL training process. Initially, a genetic algorithm randomly initializes a population with different leg length. These individuals are then trained using reinforcement learning. Rewards serve as the fitness metric for updating the population. The algorithm eventually converges to the individual with the highest fitness, resulting in optimized leg length.}
    \label{fig:example}
\end{figure}

\subsubsection{Basic Settings of RL}
In this work, we treat all structural parameters of the biped robot interaction with the environment as a Markov transfer process(MDP). The MDP is defined by a tuple $(\mathcal{S},\mathcal{A},P,r,\gamma,L)$, where $ \mathcal{S} $ denotes the state space, encompassing the possible environmental states that the robot may encounter. The action space $ \mathcal{A} $ includes the diverse actions available to the robot. The state transition probability function $ P(s_{t+1}|s_t,a_t) $ characterizes the likelihood of the system transitioning from one state to another. The reward function $ r(s_t,a_t) $ delineates the immediate reward received by the robot upon taking a specific action. The $ \gamma $ stands for the discount factor, indicating the extent to which future rewards are discounted, typically ranging between 0 and 1.  The goal is to find the optimal policy $ \pi^* $  that maximize the discounted reward:

\begin{equation}
\pi^* = \arg\max_{\pi} \mathbb{E}\left[\sum_{t=0}^{\infty} \gamma^t R_t \right]
\end{equation}

\textbf{State Space:}  
In the training phase of SERL (Structured Environment Reinforcement Learning), privileged information is utilized to facilitate the rapid adaptation of bipedal robots by quickly identifying optimal parameters. The state space $s_t$ encompasses proprioceptive information $o_t$ , velocity information
 $v_t$ and privileged information $e_t$, where the privileged observational information $e_t$ includes terrain data, body mass, and friction coefficients. Moreover, the impact of leg length on the interaction between the robot and its environment presents a challenging modeling task. So we take the structural information $L_i$ as one of the observable parameters, and both $L_i$ and $e_t$ are mapped to a latent space to generate the latent variable $z_t$. This process aims to obtain an implicit representation of the interaction between the robot’s structural parameters and the environment, enabling the robot to more rapidly adapt to changes in leg length and environmental conditions.
% Due to the limited information obtained during physical deployment, including only proprioceptive perception, a teacher-student policy is employed to enhance training efficiency and quality in the simulation environment.

% For the teacher policy, the state space $s_t$ comprises proprioceptive information $o_t$, body linear velocity $ v_t $, terrain information $h_t$, and privileged information $e_t$, such as body mass, friction, and structural dimensions. To reduce data dimensionality and extract key features, $h_t$ and $e_t$ are processed through an Env Encoder, mapping the data to a latent space and generating latent variable $z_t$, i.e.,

\begin{equation}
	s_t=o_t+z_t+v_t
\end{equation}
	
\textbf{Action Space:}  
The action $a_t$ consists of the desired positions for the nine motors corresponding to the biped robot. The joint torque is obtained by converting the desired joint positions into joint torques using a PD controller.

\textbf{Reward Design:}
The design of rewards is crucially aligned with our ultimate goal for the robot's design. We have established a comprehensive reward system for the humanoid robot's walking performance based on \cite{leggedrobot}. In addition, we added the task reward. For example, in a comprehensive locomotion task, the task reward is based on tracking a specified speed. For the achieving maximum velocity task, the task reward is based on attaining higher speeds. More details on reward settings and training can be found in the \href{SERL-IROS2024.github.io}{SERL-IROS2024.github.io} appendix here.

% Reward design is a crucial factor affecting the effectiveness of reinforcement learning training. In the task of complex terrain walking for the biped robot, the reward $r$ is divided into three main components: basic reward $r_b$, posture reward $r_p$, and task reward $r_t$, i.e.,

% \begin{table}[ht]
% \caption{Reward Design.}
% \label{Reward Design}
% \renewcommand{\arraystretch}{1.3}
% \begin{center}
% \begin{tabular}{ l l}
% \toprule
% Reward  & Equation$(r_i)$ \\
% \midrule
 
% xy-axis Linear Velocity   &  $exp\{-4\cdot(v_{xy}^{cmd}-v_{xy}^2)\}$  \\
% z-axis Angular Velocity    & $exp\{-4\cdot(\omega_{yaw}^{cmd}-\omega_{yaw}^2)\}$  \\
% xy-axis Angular Velocity  &  $v_z^2$  \\
% z-axis Linear Velocity  &  $\omega_{xy}^2 $ \\ 

% Joint Acceleration  & $\Ddot{\theta}^2$   \\
% Joint Torques &  $\tau^2$   \\
% %Joint Torques 2 &    &  -1.0e-7 \\
% %Joint Position Limits &    &  -10.0 \\
% Action Rate &  $(a_{t-1}-a_t)^2$    \\
% Smoothness &  $ (a_t-2a_{t-1}+a_{t-2})^2 $   \\

% Body Orientation &  $(0.5-0.5g_z)^2$ \\
% Body Height &  $(h^{des}-h)^2$  \\
% %Hip Position &    &  -0.4 \\
% %Foot Clearance &    &  -10.0 \\
% %Punish Flying &    &  0.2 \\

% \bottomrule
% \end{tabular}
% \end{center}
% \vspace{-20pt}
% \end{table}

\subsubsection{Method for Structural Evolution} 
\vspace{10pt}

The process of the structural evolution  begins by initializing a population of individuals, each representing a unique set of leg length. The leg length, denoted as $l_{t_i}$ and $l_{s_i}$, are randomly selected for each individual, this initialization step creates an initial set of individuals $l_i$.

Each individual within the population undergoes reinforcement learning training to assess its performance, resulting in the output of a reward value ${R}_i$. This training process focuses on optimizing the leg length based on the rewards obtained over a specified number of training episodes.

Genetic operations, encompassing crossover and mutation, are then applied to individuals to generate new sets of leg length. Crossover involves combining the leg length of two individuals, while mutation introduces minor adjustments to an individual's leg length.

The newly generated individuals from crossover and mutation undergo RL training again. The optimization process iterates between RL training and genetic operations until a predetermined number of iterations is reached or when the variance of rewards among the population falls below a defined threshold. Eventually, the structural evolution process for the bipedal robot is completed, resulting in optimized leg length and control policy.

\subsubsection{Policy Training for Physical Deployment}
After determining the optimal leg length through SERL in the first phase to obtain deployable adaptive motion control policy based on proprioceptive sensing, we distilled the RL policy from the first phase online. In the second phase, interactions are exclusively with robots using the optimal leg length. The state space $s_t$ combines proprioceptive information $o_t$ and estimated body linear velocity $\hat{v_t}$. Unlike traditional high-dimensional state processing methods that use CNNs for dimensionality reduction\cite{RMA}, this study employs GRUs to transform high-dimensional historical state information into low-dimensional vectors $y_t$, showcasing GRUs' superior temporal sequence capturing ability compared to CNNs, the specific framework is shown in the Figure 3. The state space is defined as:
\begin{equation}
	s_t=o_t+\hat{v_t}+y_t 
\end{equation}

The adaptive motion policy based on the optimal leg length is iteratively trained under the supervision of the phase one policy.

% To implement the student policy for physical deployment, the state space $s_t$ encompasses proprioceptive information $o_t$, body linear velocity $ \hat{v_t}$ which is estimated from $o_t$ . Conventional approaches to handle high-dimensional historical state information often involve the use of CNN \cite{RMA} to reduce input dimensionality. In contrast, we employ GRU to transform the high-dimensional historical states into a low-dimensional vector $y_t$, capitalizing on GRU's superior ability to capture temporal relationships compared to CNN. Thus, the state space can be defined as:  $$ s_t=o_t+\hat{v_t} +y_t $$

% Following this, under the supervision of the teacher policy, the student policy generates the action $a_t$ applied to the physical robot,  which also consists of the desired positions for the nine motors.

\begin{figure}[h]
    \centering
    \includegraphics[width=\columnwidth]{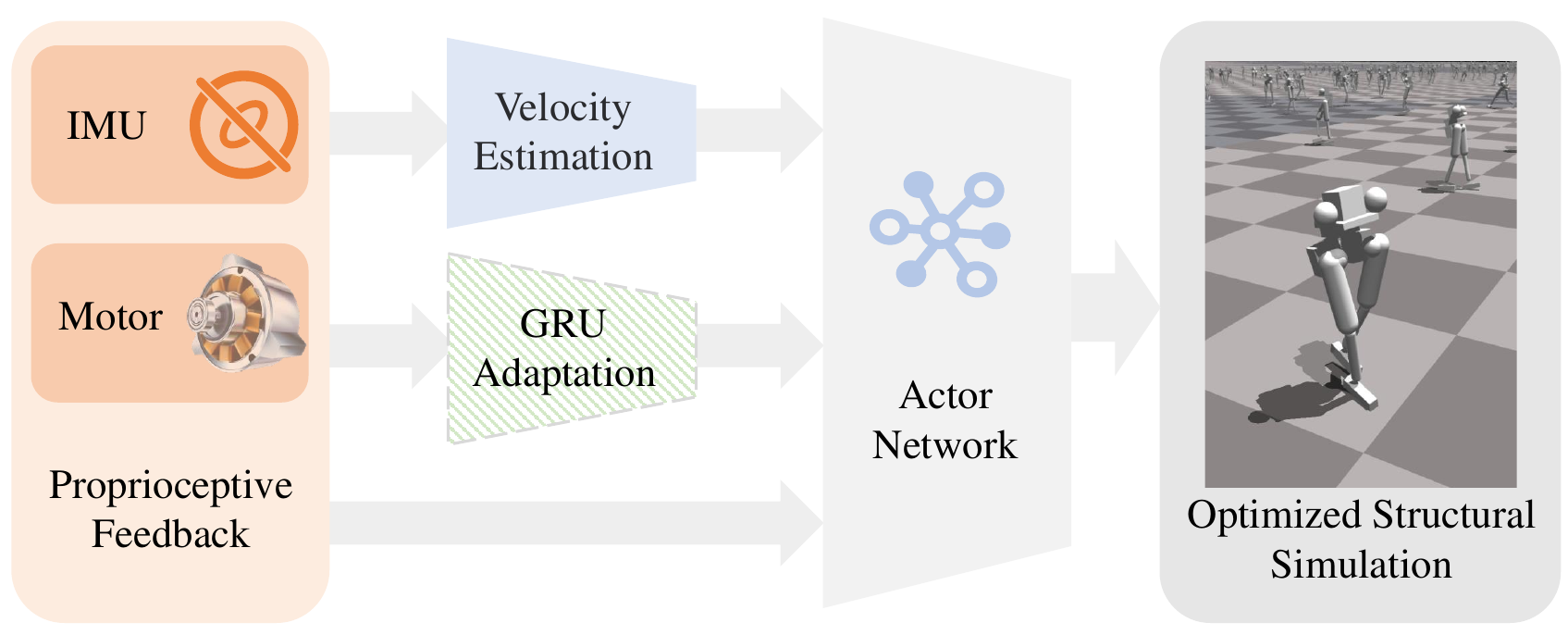}
    \caption{Overview of adaptation framework. Figure 3 shows the second stage of training. We use the leg length results and control policy obtained from the first stage of the SERL algorithm. The control policy is distilled to obtain a control policy that can be applied to a real robot.}
    \label{fig:example}
\end{figure}

\subsubsection{Train Details}

As shown in the pseudocode \ref{algorithm}, the algorithm initializes the weight parameters of the neural network and the population of the genetic algorithm. It randomly selects different combinations of leg length $L=(l_t, l_s)$ for each individual, with the population size set to $pop\_size=250 $. Subsequently, each individual undergoes training in a simulation environment, aiming to maximize the reward obtained for completing locomotion tasks. After training, the algorithm evaluates the training results for each individual and selects the optimal individuals for crossover, mutation, and selection, generating a subpopulation of individuals. This subpopulation iteratively repeats the aforementioned process until the specified number of cycles is reached.  Ultimately, the algorithm outputs optimal individual parameters, achieving leg joint length and robot motion control policy optimization.

We initially load robot models with varying pole length on the Isaac Gym platform and conduct walking training using a unified reward mechanism. To comply with physical laws, we set the unit of leg length to have a uniform mass density. Changes in length will result in an increase in both mass and moment of inertia. After completing 500 iterations of training, we further apply the SERL algorithm for optimization iterations. This step is designed to prevent the optimization goal of the SERL algorithm from biasing towards the combination of pole length that learns the fastest, neglecting the optimal combination for a specific task. In the application of the SERL algorithm, adjusting hyperparameters ensures the convergence speed of the policy network and the genetic algorithm is consistent, thus avoiding falling into local optima.

Moreover, the concept of "fair rules" is introduced during the training process. The optimization goal of the genetic algorithm is the combination of pole length that achieves the maximum reward under the same task conditions. However, different command resampling, random thrust sizes, and other factors may introduce unfair criteria for evaluation. Therefore, it is necessary to ensure these parameters are consistent for each individual to guarantee the fairness of the assessment.The specific hyperparameters can be referred to in Table \ref{Hyperparameters}.

In the first phase of the algorithm, SERL, we use a single NVIDIA RTX-4090 GPU to train on the Isaac Gym simulation platform, which takes approximately 12 hours. In the second phase, we use the same hardware configuration as in the first phase, with a training time of about 1 hour.

\begin{table}[h]
\caption{Hyperparameters of Structure Evolution.}
\label{Hyperparameters}
\renewcommand{\arraystretch}{1.3}
\begin{center}
\begin{tabular}{ l r}
\toprule
Parameter& Value \\
\midrule
Population Size& 250 \\

DNA Length $(bit)$ & 9 \\

Thigh Length Range $(m)$ & (0.2,0.4) \\

Shin Length Range $(m)$ & (0.2,0.4) \\

Resolution $(m)$ & 0.01 \\

Crossover Rate  & 0.8 \\

Mutation Rate  & 0.03 \\

Steps per Iteration & 96\\

Iterations per Evolution & 10\\

Interval of Pushing Robots $(s)$ & 5\\
\bottomrule

\end{tabular}
\end{center}
\vspace{-20pt}

\end{table}

\begin{algorithm}
    \caption{Structure Evolution via Reinforcement Learning}
    \SetAlgoNlRelativeSize{0}
    \label{algorithm}
    \SetKwInput{KwIn}{Input}
    \SetKwInput{KwOut}{Output}
    
    \KwIn{thigh and shin length range $(l_{\text{t}}^{\min },l_{\text{t}}^{\max })$,   $(l_{\text{s}}^{\min },l_{\text{s}}^{\max })$,  evolution iteration $N_e$, policy iteration per evolution iteration $N_{it}$, Population $P$,  index of individual $i$, selection probability $p_{s(i)}$, crossover probability $p_{c}$, mutation probability $p_{m}$.}

    \KwOut{optimal $l_{\text{t}}$, $l_{\text{s}}$, $reward$.}
    
    Generate the initial population and encode\;
    \For{$i_e$ in [1...$N_e$]}{
        Reset morphology of the robot ($P_{i_e}[l_{\text{t}}^{\min },l_{\text{t}}^{\max }]$), ($P_{i_e}[l_{\text{s}}^{\min },l_{\text{s}}^{\max }]$) and physical model (collision, mass, inertia)\;
        \For{$i_t$ in [1...$N_{it}$]}{
            \For{$k$ in [1...$N_{\text{steps}}$]}{
                Get $Act_{k}$ from policy\;
                Parallel step simulation $\to$ $Obs_{k}$, $Rew_{k}$\;
            }
            $Rew(i_t) = [Rew_{1}...Rew_k]$\;
            $Obs(i_t) = [Obs_{1}...Obs_k]$\;
            Policy update\;
        }
        $Rew(i_e) = [Rew_{1}...Rew_{i_t}]$\;
        $Obs(i_e) = [Obs_{1}...Obs_{i_t}]$\;
        $fitness = \sum Rew(i_e) - \min(\sum Rew(i_e) + \epsilon)$\;
        Select individuals using roulette wheel selection: $p_{s(i)} = \frac{fitness(i)}{\sum fitness}$\;
        
        $p(i_e+1) \leftarrow$ Crossover and mutation $\leftarrow p_{c} \cdot p_{m}$\;
    }
\end{algorithm}

%*********************************************************************
\subsection{Bipedal Robot Platform Design} 
\subsubsection{Overview}
We introduce Wow Orin, an innovative bipedal robot featuring nine degrees of freedom, encompassing hip joints for internal/external rotation, hip flexion/extension, ankle flexion/extension, knee flexion/extension in each leg, and the abduction/adduction joints of the hip are driven by a single shared motor. All joints are powered by Unitree A1 motors with a peak torque of $33.5N\cdot m$ and a maximum speed of $21 rad/s$. Wow Orin is designed with a strong emphasis on lightweight principles, weighing a mere $10.5kg$ and standing at a height of $0.88m$. The thigh and shin dimensions are optimized using the SERL algorithm, resulting in values of $l_t=0.31 m$ and $l_s=0.36 m$ respectively. To reduce leg rotational inertia and enhance motion performance, the knee joint incorporates a synchronou belt drive, while the ankle joint utilizes a Bowden cable drive. Wow Orin is equipped with a Jetson Orin NX (8GB) onboard PC, delivering an impressive AI performance of $40 TOPS$.

\begin{figure}[b]
    \centering
    \includegraphics[width=\columnwidth]{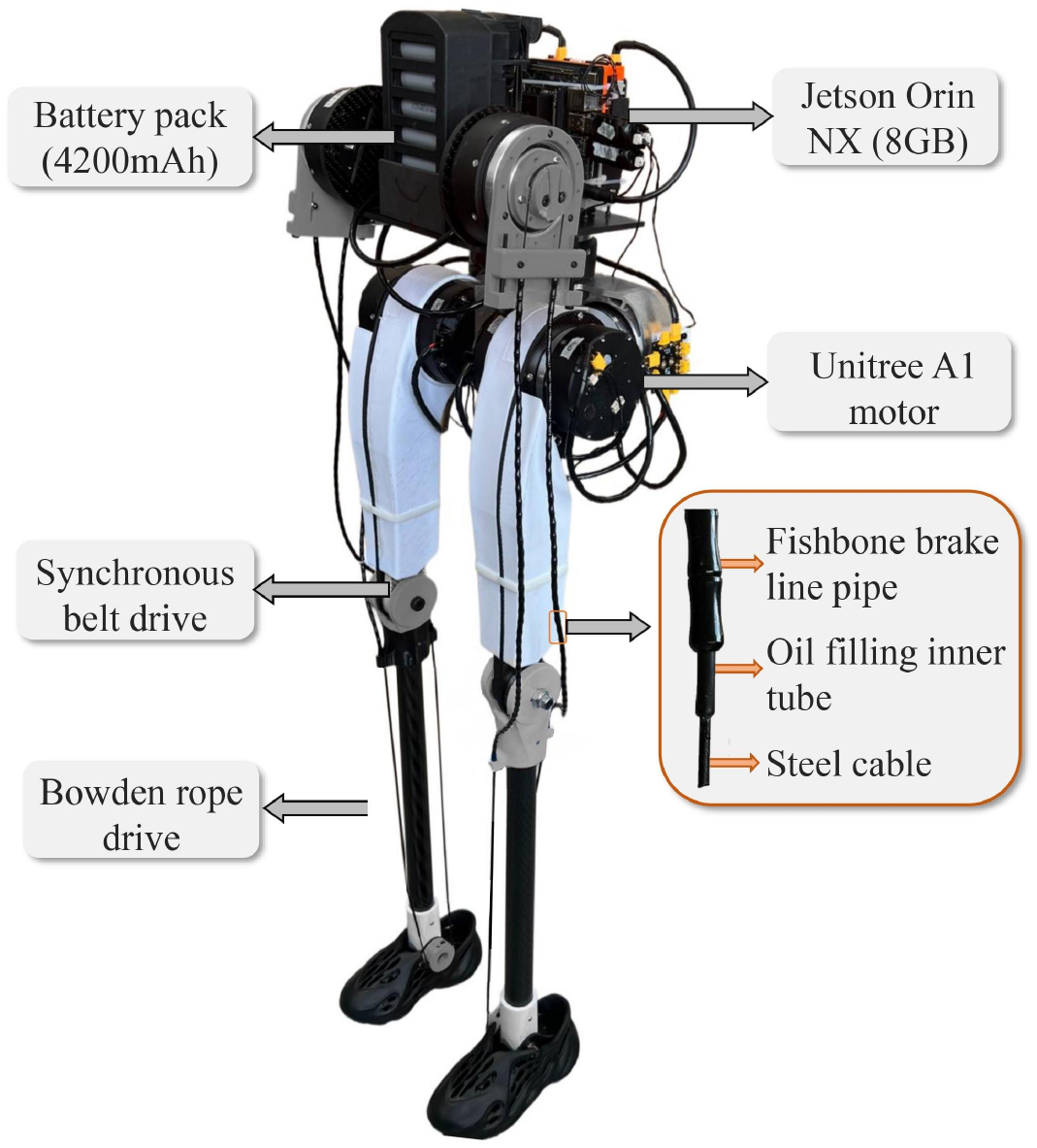}
    \caption{Robot structure design.}
    \label{Robot structure design}
\end{figure}

\subsubsection{Lightweight Leg Design}
To reduce the rotational inertia of the legs and further enhance motion performance, besides the aforementioned ankle joint rope drive method, the knee joint adopts a synchronous belt drive. In addition, we used carbon fiber as the skeleton for the robot's lower and upper leg links, 3D printed PETG material for some connectors, and aluminum alloy for key load-bearing parts to achieve the ultimate reduction in weight without sacrificing structural strength. Through the above design, the weight of the robot's single leg was reduced to 1.3kg. We designed the joint parameters as follows in the Table \ref{Joint parameters}.

\newcommand{\tabincell}[2]{\begin{tabular}{@{}#1@{}}#2\end{tabular}}
\begin{table}[h]
\caption{Joint parameters}
\label{Joint parameters}
\renewcommand{\arraystretch}{1.3}
\begin{center}
\begin{tabular}{ l c c c r}
\toprule
Name& \tabincell{c}{Torque\\ $(N\cdot m)$ } & \tabincell{c}{Limit\\ $(rad)$} & \tabincell{c}{Max speed \\$(rad/s)$} & Drive mode \\
\midrule
Hip yaw & $33.5$ & $(-0.1,1)$ & $21$ & Direct\\
Hip roll & $33.5$ & $(-0.2,1.5)$ & $21$ & Direct\\
Hip pitch & $33.5$ & $(-1.5,1.5)$ & $21$ & Direct\\
Knee     & $50.25$ & $(-1.5,1.5)$ & $14$ & Belt\\
Ankle    & $67$ & $(-0.75,0.75)$ & $10.5$ & Bowden cable\\
\bottomrule
\end{tabular}
\end{center}
\end{table}

\subsubsection{Bionic Fishbone Bowden Cable Driven Ankle Joint}
The design of the ankle joint is our main distinction from other robots, which use smaller mass motors in the ankle joint to reduce leg inertia, resulting in the maximum force generated by the ankle joint being significantly less than that of the thigh joint. According to research in literature, the torque generated by the human ankle joint during running is nearly equal to that of the thigh joint. Therefore, based on previous work, following the principle of mass concentration, and to meet the requirements of mass elevation in structural design, we adopted a Bowden cable to implement a series elastic actuation system, allowing the ankle actuator to be positioned anywhere outside the joint itself. This eliminates the need to consider the mass of the ankle joint motor affecting leg movement, significantly reducing leg inertia mass while greatly enhancing the torque of the ankle joint and significantly increasing the overall structural design flexibility. On our robot, the ankle joint power motor is placed in the upper body and transmits power to the ankle joint through the Bowden cable.

We adopted a bionic fishbone structure for the Bowden cable (as shown in the Figure \ref{Robot structure design}), which uses multiple segments closely connected by spherical joints to achieve structural flexibility and adjustability. By adjusting the number of segments, the total length of the structure can be rapidly changed, thus providing high adaptability. The interlayer part is made of a complete oil tube filled with lubricant, significantly reducing the friction encountered by the steel wire during movement. Compared to traditional Bowden cable structures, this design not only significantly improves load-bearing capacity but also achieves extremely low friction.

%*********************************************************************

%*********************************************************************

\section{EXPERIMENTAL EVALUATIONS}
\subsection{Evaluation of the SERL Algorithm}
\subsubsection{Effectiveness of the Algorithm Optimization Process}

In order to demonstrate the effectiveness of SERL algorithm in enhancing robot motion performance, we first define two different task forms as specific design objectives: comprehensive locomotion under perturbations and achieving maximum velocity. The two different tasks are represented by different task reward forms, denoted as $r_t$. Through data analysis of the mean reward of individuals in the population and the maximum fitness in the population for both tasks, where population fitness is used to measure the maximum difference in individual performance within
the population and can effectively demonstrate the convergence of the population, we can effectively showcase the convergence effect of the population. We also record the thigh and shin length during the population evolution process. The convergence process of the above data reflects the improvement of policy and performance and the stable variation of design parameters.
\begin{figure}[h]
    \centering
    \includegraphics[width=.4\textwidth]{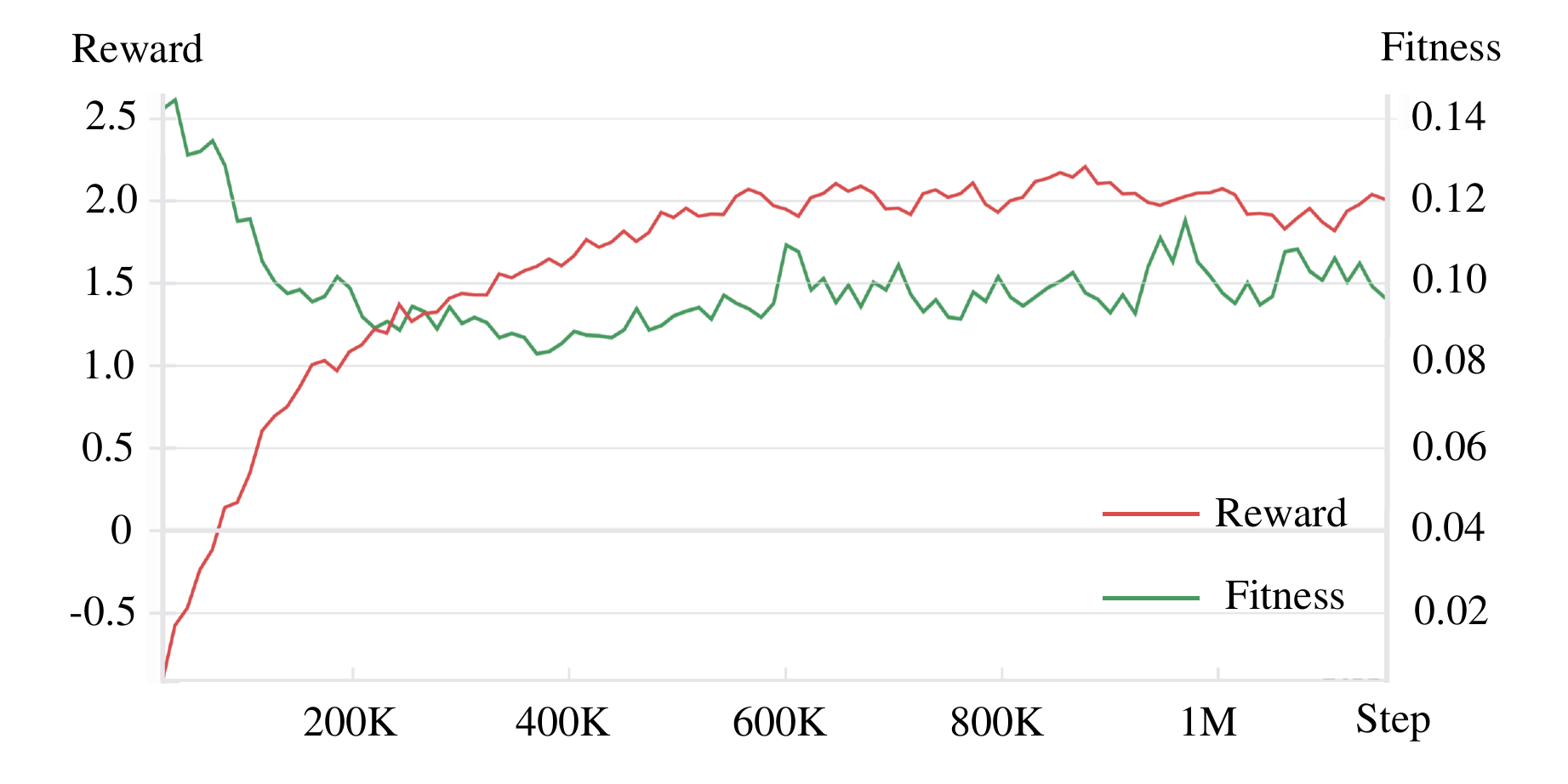}
    \caption{Reward and maximum fitness of comprehensive locomotion task.}
    \label{Reward and maximum fitness of comprehensive locomotion task}
    \vspace{-20pt}
\end{figure}

\begin{figure}[h]
    \centering
    \includegraphics[width=.4\textwidth]{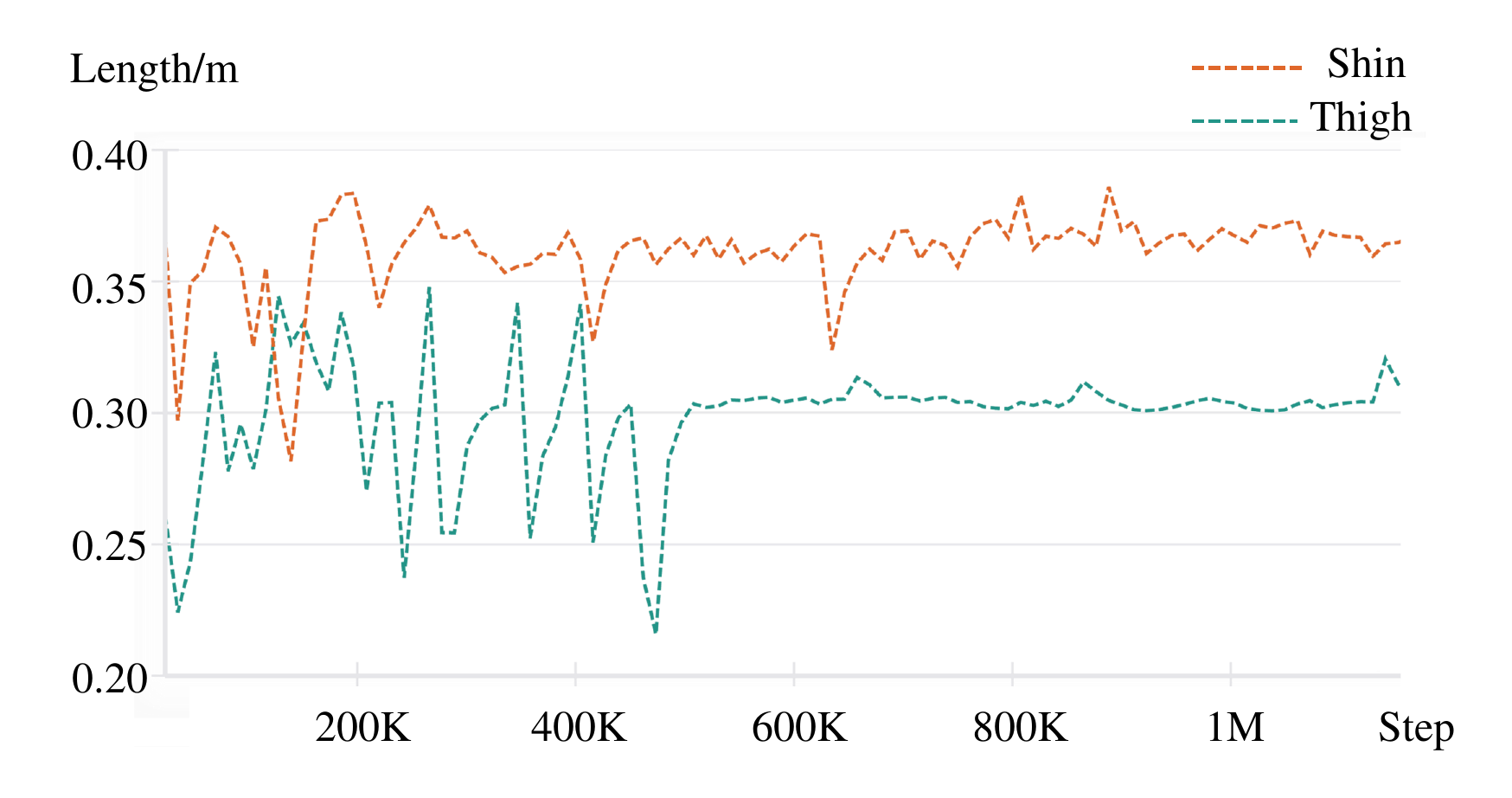}
    \caption{Thigh and shin length of comprehensive locomotion task.}
    \label{Thigh and shin length of comprehensive locomotion task}
    \vspace{-20pt}
\end{figure}

\begin{figure}[h]
    \centering
    \includegraphics[width=\columnwidth]{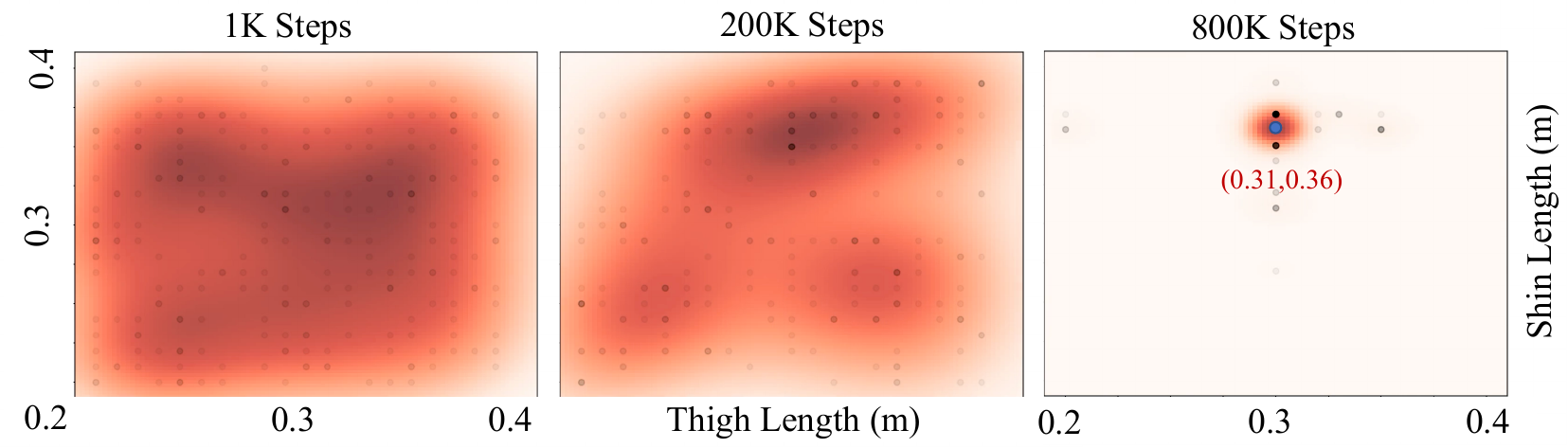}
    \caption{Thigh and shin length distribution of comprehensive locomotion task.}
    \label{Thigh and shin length distribution of comprehensive locomotion task}
    \vspace{-10pt}
\end{figure}

For the comprehensive locomotion under perturbations task, as the results shown in Figures \ref{Reward and maximum fitness of comprehensive locomotion task}, ref \label{Thigh and shin length of comprehensive locomotion task} and  \ref{Thigh and shin length distribution of comprehensive locomotion task}, the leg length parameters converge to their optimal values at around 700k steps, with the population's mean reward reaching its maximum and the population's maximum fitness reaching its minimum. The population distribution of leg length also changed from scattered to gradually concentrated. Surprisingly, the convergence results show that longer shin length compared to thigh length lead to better task completion. This finding is different from previous methods based on biological inspiration or manual adjustment, where thigh and shin length are typically set to the same value in bipedal robot design.

% \begin{figure}[t]
%     \centering
%     \includegraphics[width=\columnwidth]{reward.pdf}
%     \caption{Mean reward and maximum fitness of tracking velocity command task}
%     \label{fig:example}
% \end{figure}
% \begin{figure}[h]
%     \centering
%     \includegraphics[width=\columnwidth]{leg_length.pdf}
%     \caption{Thigh and shin length of tracking velocity command task}
%     \label{fig:example}
% \end{figure}

\begin{figure}[h]
    \centering
    \includegraphics[width=.4\textwidth]{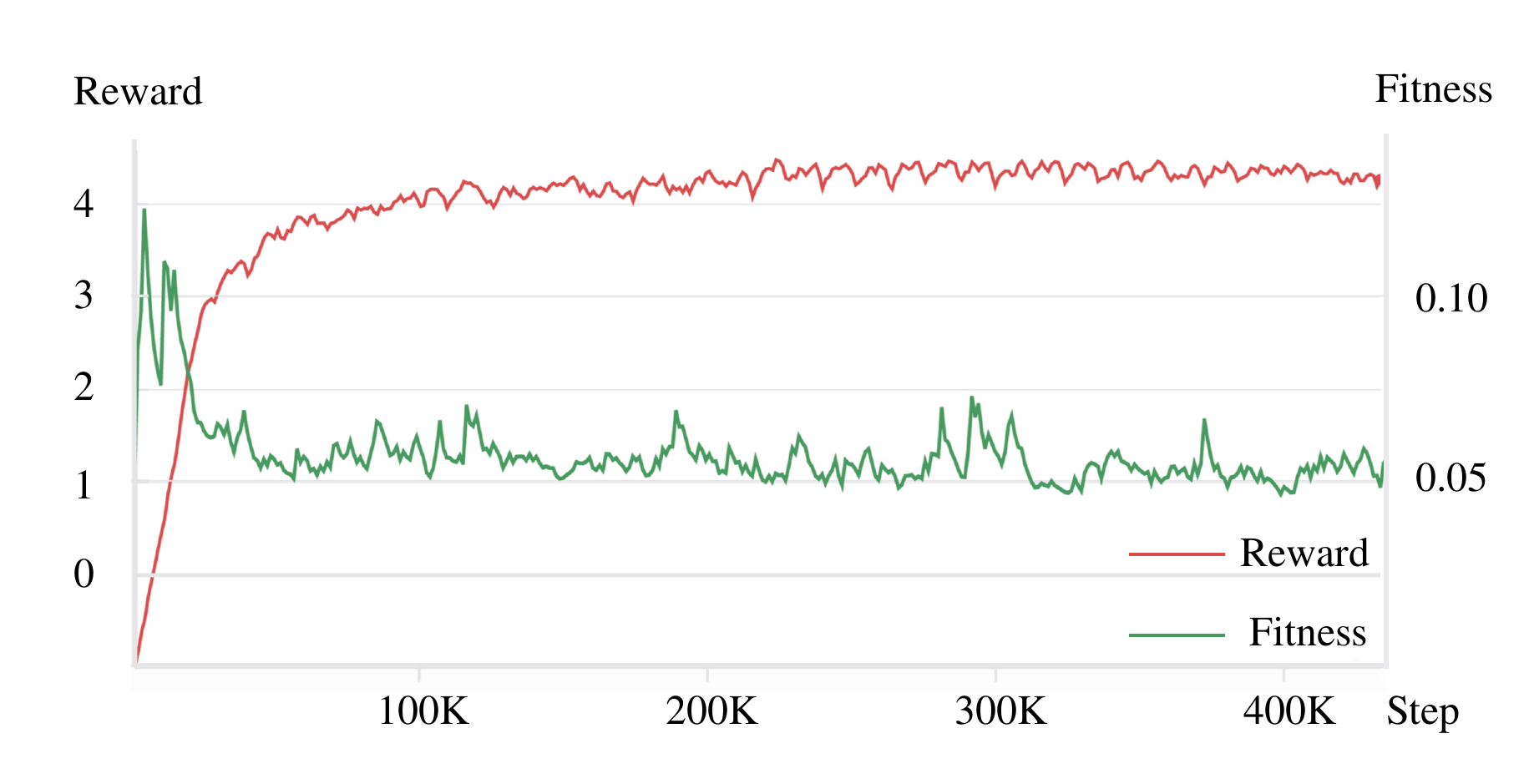}
    \caption{Reward and maximum fitness of achieving maximum velocity.}
    \label{Reward_and_maximum_fitness_of_achieving_maximum_velocity}
    \vspace{-10pt}
\end{figure}

\begin{figure}[h]
    \centering
    \includegraphics[width=.4\textwidth]{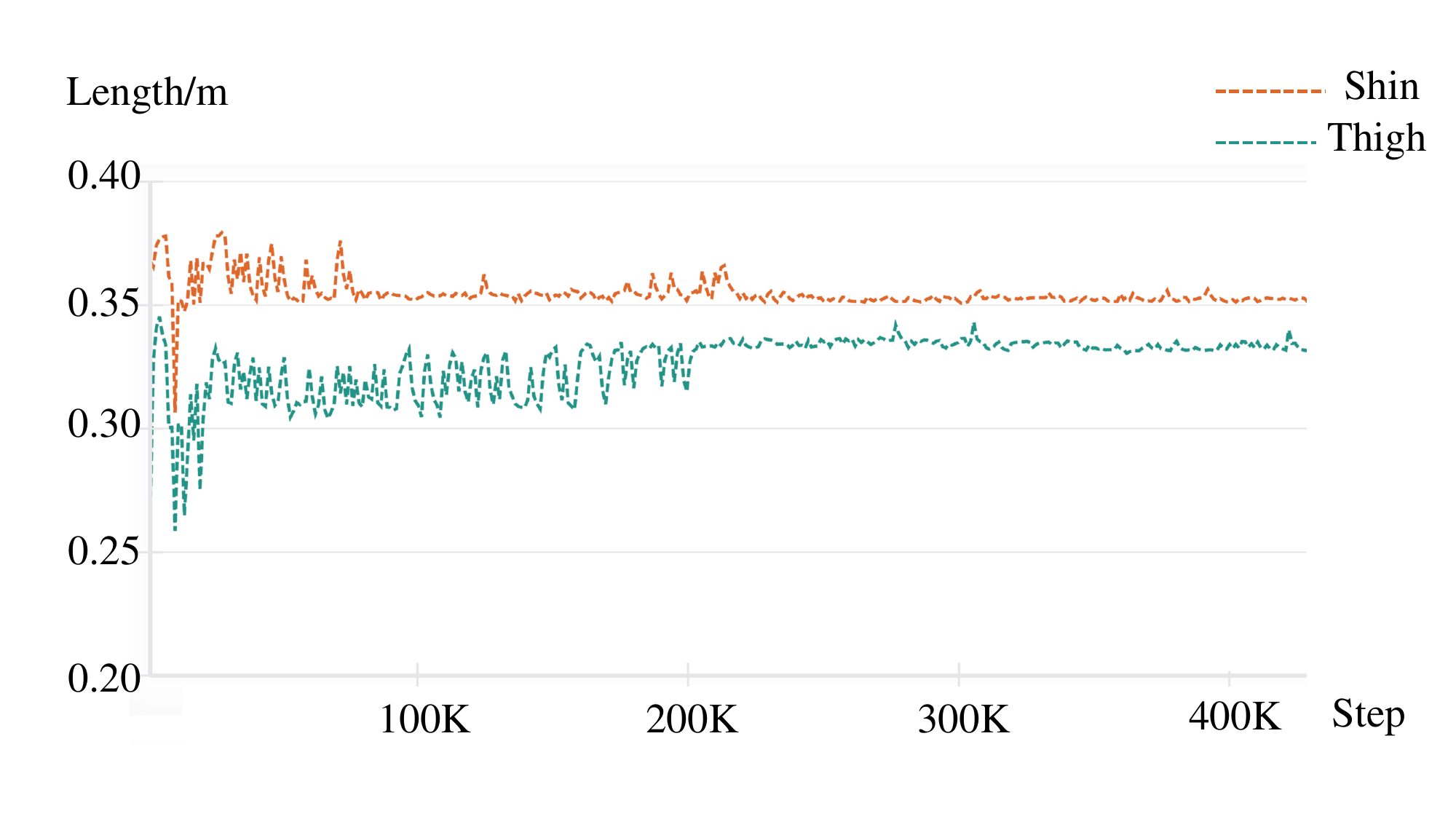}
    \caption{Thigh and shin length of achieving maximum velocity task.}
    \label{Thigh and shin length of achieving maximum velocity task}
    \vspace{-20pt}
\end{figure}

\begin{figure}[h]
    \centering
    \includegraphics[width=\columnwidth]{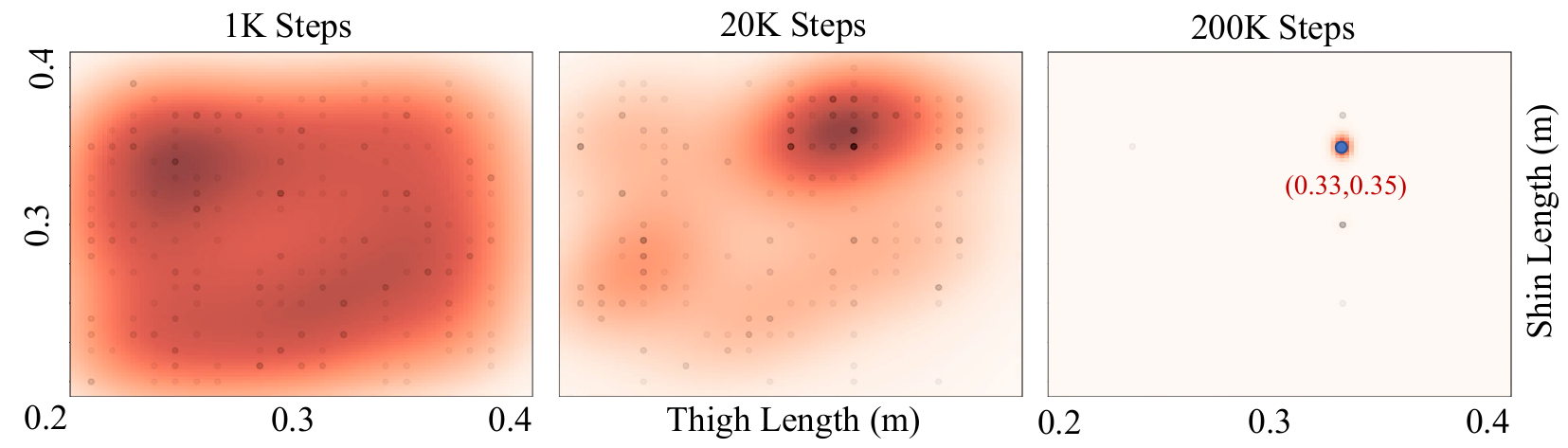}
    \caption{Thigh and shin length distribution of achieving maximum velocity task.}
    \label{Thigh and shin length distribution of achieving maximum velocity task}
    \vspace{-10pt}
\end{figure}

In the results of the achieving maximum velocity task shown in Figures \ref{Reward_and_maximum_fitness_of_achieving_maximum_velocity}, \ref{Thigh and shin length of achieving maximum velocity task} and \ref{Thigh and shin length distribution of achieving maximum velocity task}, we found that the convergence is faster, reaching a good convergence effect after 200k steps. The results show that, compared to the comprehensive locomotion task, in the task of achieving maximum velocity, shin length increase while thigh length decrease slightly, and the length of the thigh and shin tend to be the same.

The convergence results of the two tasks indicate that for different task forms, the SERL algorithm can effectively provide different optimal design solutions, satisfying diverse requirements in bipedal robot design and maximizing robot performance and capabilities.

%test
% \begin{figure}[pt]
%     \centering
%     \includegraphics[width=\columnwidth]{reward.pdf}
%     \caption{Mean reward and maximum fitness of tracking velocity command task1}
%     \label{fig:example}

%     \centering
%     \includegraphics[width=\columnwidth]{leg_length.pdf}
%     \caption{Thigh and shin length of tracking velocity command task1}
%     \label{fig:example}

%     \centering
%     \includegraphics[width=\columnwidth]{speed_reward.pdf}
%     \caption{Mean Reward and maximum fitness of achieving maximum velocity1}
%     \label{fig:example}

%     \centering
%     \includegraphics[width=\columnwidth]{speed_leg_length.pdf}
%     \caption{Thigh and shin length of achieving maximum velocity1}
%     \label{fig:example}

% \end{figure}

% \\
% \\
\begin{figure}[h]
    \centering
    \includegraphics[width=.4\textwidth]{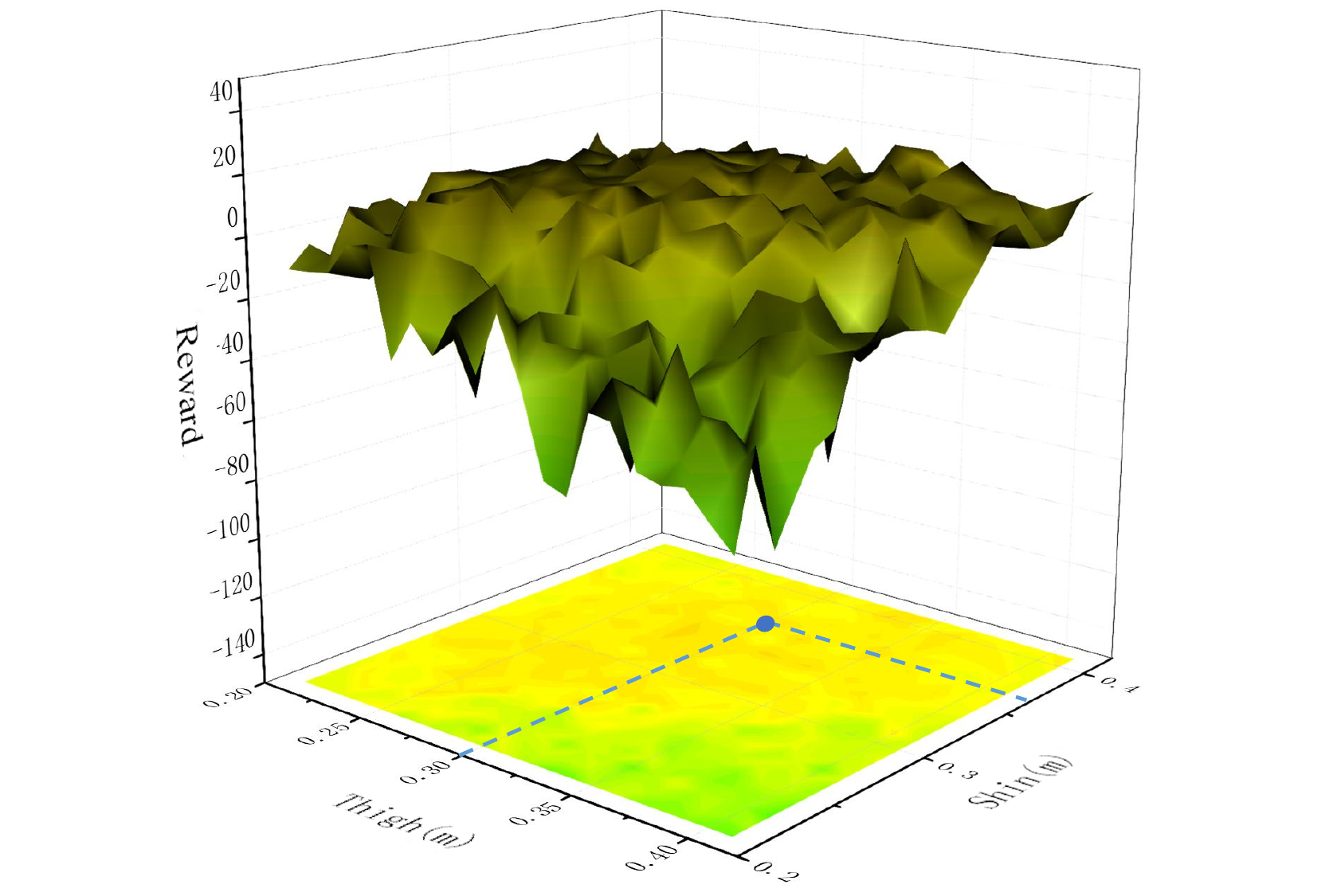}
    \caption{Rewards within the design space of comprehensive locomotion task.}
    \label{base3D}
    \vspace{-15pt}
\end{figure}

\begin{figure}[h]
    \centering
    \includegraphics[width=.4\textwidth]{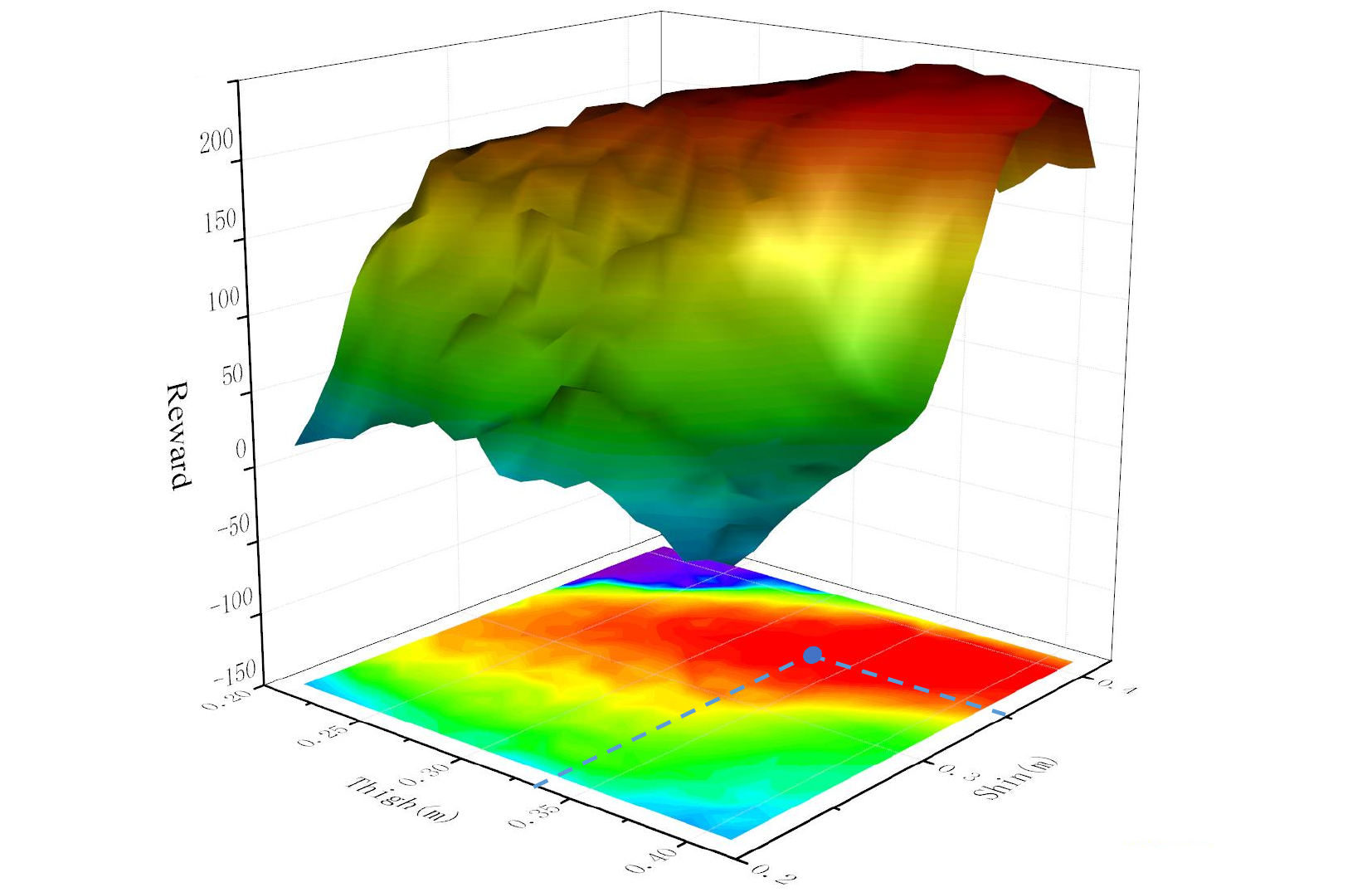}
    \caption{Rewards within the design space of achieving maximum velocity task.}
    \label{speed3D}
    \vspace{-20pt}
\end{figure}
\subsubsection{ Compare with other parameters within the design space}

We compare the optimal leg length obtained using the SERL algorithm with other leg length within the design space. As shown in Figures \ref{base3D} and \ref{speed3D}, the experimental results clearly demonstrate that the manually specified leg length yields optimal results consistent with the SERL algorithm optimization results, thus validating the effectiveness of the SERL algorithm. Additionally, we observe that for the comprehensive locomotion task, the optimal reward corresponds to a wider range of pole length, indicating that the algorithm requires more time to converge during the training process. In contrast, for the maximum velocity task, the distribution range of the global optimal solution is smaller, suggesting faster convergence of the algorithm. This further confirms the adaptability of the algorithm to real-world scenarios.

% \small
% \begin{table}[h]
% \caption{Hyperparameters}
% \label{table_example}
% \renewcommand{\arraystretch}{1.3}
% \begin{center}
% \begin{tabular}{ l r}
% \toprule
% Parameter& Value \\
% \midrule
% Population Size& 250 \\

% DNA Length $(bit)$ & 9 \\

% Thigh Length Range $(m)$ & (0.2,0.4) \\

% Shin Length Range $(m)$ & (0.2,0.4) \\

% Resolution $(m)$ & 0.01 \\

% Crossover Rate  & 0.8 \\

% Mutation Rate  & 0.03 \\

% Steps er Iteration & 96\\

% Iterations per Evolution & 10\\

% Interval of Pushing Robots $(s)$ & 5\\
% \bottomrule

% \end{tabular}
% \end{center}
% \end{table}

\subsection{Performance Testing of Wow Orin Robot}

To assess the superior performance of Wow Orin and demonstrate its high energy efficiency and agility, we employed the same algorithmic settings to train the control policy of Wow Orin, Cassie, and Unitree H1 in a flat simulation environment, followed by comparative tests of their Cost of Transport (COT), maximum velocity and Froude number\cite{Froude}. 
\subsubsection{Energy  Efficiency Comparison}
The COT is a common metric used to gauge the efficiency of legged animals and robots, calculated as $COT = P/(m·g·v)$, where $P$ denotes joint power, $m$ represents the robot's weight, $g$ is the acceleration due to gravity, and $v$ is the robot's forward velocity.As shown in Table \ref{COT}, through extreme lightweight design, Wow Orin achieves the lowest weight among the three robots, at only 10.5 kg. Additionally, it employs motors matched to the body mass for propulsion, resulting in minimal joint torque. The COT results demonstrate that Wow Orin significantly outperforms the other two robots, indicating its superior energy efficiency. This allows the robot to achieve longer endurance with limited battery capacity, which is crucial for tasks requiring prolonged movement. Since the calculation process of COT excludes the influence of the robot's weight, it indicates to some extent that the optimization of our robot's leg length design plays a crucial role in achieving a lower COT.

% \subsubsection{Performance in Harsh Environments}
% To test the adaptability of Wow Orin in complex terrains, we trained the control policy of three bipedal robots using the same environment and reward settings. Subsequently, we employed different terrain configurations, including slopes, stairs, and trimesh terrains (see attached terrain settings). The robots traversed the aforementioned terrains at specified speeds, and we recorded both the success rate of the robots' traversal and the tracking error of their speeds. The experiment results are presented in the table. The results indicate that Wow Orin demonstrates superior terrain adaptability across the three different terrain tests.

\subsubsection{Maximum Velocity on Flat Ground}
The Froude number is a dimensionless quantity used to compare the dynamic similarity of different-sized robots or organisms under similar motion conditions\cite{Froude}, especially in walking or running scenarios. Its calculation formula is $F_r = v^2 / {gl}$, where $v$ is the speed of the robot's movement, $g$ is the acceleration due to gravity, and $l$ is the length of the robot's legs. We trained the three robots with the same parameters and environmental settings to achieve their maximum speed on flat ground. The results, as shown in the Table \ref{COT} , reveal that Wow Orin's maximum speed and Froude number exceeds that of Unitree H1 but falls short of Cassie's top speed. This difference may be attributed to Cassie's more explosive leg structure and the lower peak torque of the joint motors used in Wow Orin compared to Cassie's.

\begin{table}[h]
\caption{Performance Testing of Wow Orin Robot.}
\vspace{-10pt}
\label{COT}
\renewcommand{\arraystretch}{1.3}
\begin{center}
\begin{tabular}{ l c c c}
\toprule
  & Wow Orin & Cassie & Unitree H1 \\
\midrule
Weight $(kg)$ & 10.5 & 31 & 47 \\
Joint Maximum Torque  $(N\cdot m)$ & 67 & 195 & 360 \\
Cost of Transport ( COT ) & \textbf{0.407} & 0.762 & 0.718 \\

\hline
Leg Length$(m)$ & 0.7 & 1.0 & 0.8 \\
Max Velocity $(m/s)$ & 2.1 & 3.2 & 1.9 \\
Froude number & 0.75 & 1.04 & 0.46 \\

\bottomrule
\end{tabular}
\end{center}
\vspace{-10pt}
\end{table}
\section{CONCLUSIONS}

This study demonstrates the effectiveness of the SERL algorithm in the structural parameter design of bipedal robots, offering valuable insights for advancing this field. By combining reinforcement learning motion control policy with evolutionary algorithms, the SERL algorithm successfully identifies structural parameter values within the specified design space that best meet task requirements. Through comparison with other parameters within the design space, we validate the exceptional performance of the SERL algorithm in achieving theoretically optimal values. Furthermore, we apply the optimized results of the SERL algorithm to practical design, creatively developing the Wow Orin bipedal robot. 
In this paper, we primarily focus on optimizing the leg length as a starting point for the design. In the future, we will apply the SERL algorithm to other parameters and tasks, and improve the algorithm to achieve the co-optimization of multiple design parameters.

\addtolength{\textheight}{-12cm}   % This command serves to balance the column length
                                  % on the last page of the document manually. It shortens
                                  % the textheight of the last page by a suitable amount.
                                  % This command does not take effect until the next page
                                  % so it should come on the page before the last. Make
                                  % sure that you do not shorten the textheight too much.

%%%%%%%%%%%%%%%%%%%%%%%%%%%%%%%%%%%%%%%%%%%%%%%%%%%%%%%%%%%%%%%%%%%%%%%%%%%%%%%%

%%%%%%%%%%%%%%%%%%%%%%%%%%%%%%%%%%%%%%%%%%%%%%%%%%%%%%%%%%%%%%%%%%%%%%%%%%%%%%%%

%%%%%%%%%%%%%%%%%%%%%%%%%%%%%%%%%%%%%%%%%%%%%%%%%%%%%%%%%%%%%%%%%%%%%%%%%%%%%%%%
% \section*{APPENDIX}

% Appendixes should appear before the acknowledgment.

% \section*{ACKNOWLEDGMENT}

% The preferred spelling of the word ÒacknowledgmentÓ in America is without an ÒeÓ after the ÒgÓ. Avoid the stilted expression, ÒOne of us (R. B. G.) thanks . . .Ó  Instead, try ÒR. B. G. thanksÓ. Put sponsor acknowledgments in the unnumbered footnote on the first page.

%%%%%%%%%%%%%%%%%%%%%%%%%%%%%%%%%%%%%%%%%%%%%%%%%%%%%%%%%%%%%%%%%%%%%%%%%%%%%%%%

\end{document}